\documentclass[onecolumn,12pt,aps,epsfig,showpacs,preprintnumbers,amsmath,amssymb]{revtex4}
\usepackage{epsfig} 
\usepackage{graphicx}  
\usepackage{dcolumn}  
\usepackage{bm}
\begin{document}  
  
\preprint{}  
  
\title{Observation of Positive-Parity Bands in $^{109}$Pd and $^{111}$Pd: Enhanced $\gamma$-Softness}  

\author{E.A.~Stefanova$^1$\footnote{E-mail address: elenas@inrne.bas.bg}, S.~Lalkovski$^2$, A. Korichi$^3$, T. Kutsarova$^1$, 
A.~Lopez-Martens$^3$, F.R. Xu$^4$, H.L. Liu$^4$, S.~Kisyov$^2$, A.~Minkova$^2$, D.~Bazzacco$^5$, M.~Bergstr\"om$^6$, A.~G\"orgen$^7$\footnote{Present address: DAPNIA/SPhN, CEA-Saclay, Gif-sur-Yvette, France}, F.~Hannachi$^3$\footnote{Present address: Centre d'Etudes Nuclaires de Bordeaux-Gradignan, Gradignan, France},  B.~Herskind$^6$, H.~H\"ubel$^7$, A.~Jansen$^7$, T.~L.~Khoo$^8$, Zs.~Podoly\'ak$^9$\footnote{Present address: Department of Physics, University of Surrey, Guildford, GU27XH, UK}, and G.~Sch\"onwasser$^7$}

\affiliation{
$^1$Institute for Nuclear Research and Nuclear Energy, Bulgarian Academy of Science, 1784 Sofia, Bulgaria \\
$^2$Faculty of Physics, University of Sofia "St.~Kliment Ohridski", 1164 Sofia, Bulgaria; \\
$^3$CSNSM Orsay, IN2P3/CNRS, F-91405, France\\ 
$^4$School of Physics, Peking University, Beijing 100871, China\\
$^5$Dipartimento di Fizica e UNFN, Sezione di Padova, 1-35131 Padova, Italy\\
$^6$The Niels Bohr Institut, Blegdamsvej 17, DK-2100 Copenhagen, Denmark\\
$^7$ISKP, Universit\"at Bonn, Nussallee 14-16, D-53115, Germany\\
$^8$Physics Division, Argonne National Laboratory, Argonne, Illinois 60439, USA\\
$^9$INFN, Laboratori Nationali di Legnaro, Italy\\
}

\date{\today}  
  
  
\begin{abstract}  

The neutron-rich nuclei $^{109}$Pd and $^{111}$Pd were produced as fission fragments following the 
$^{30}$Si + $^{168}$Er reaction at 142 MeV.  
Using the identification based on the coincidences with the complementary fission fragments, the only positive-parity bands 
observed so far in $^{109}$Pd and $^{111}$Pd emerged from this work. 
A band, built on top of the 5/2$^+$ ground state exhibiting $\Delta I$ = 1 energy-level staggering, was observed in each of these
nuclei. Both nuclei of interest, $^{109}$Pd and $^{111}$Pd, are suggested to lie in the transitional region of Pd isotopes of maximum 
$\gamma$-softness. The ground states of both nuclei are predicted by TRS calculations to be extremely $\gamma$-soft with shallow 
triaxial minima. The first crossing in the new bands is proposed to be due to an alignment of $h^2_{11/2}$ neutrons.

\end{abstract}  
  
\maketitle  
  
\section{Introduction}  

The investigated in the present study $^{109}$Pd and $^{111}$Pd belong to the region of the neutron-rich nuclei with 
$A \approx$ 100 (38,40 $\le$ $Z \le$ 50, $N >$ 56). The nuclei in this region are known to exhibit a variety of structural phenomena 
and shapes and were for example a subject of detailed theoretical study by the Nilsson-Strutinski calculations with cranked Woods-Saxon
average potential in Ref. \cite{ska97}. The neutron-rich nuclei in this region undergo
transition from nearly spherical vibrational to deformed rotational character. This transition is known to be more abrupt
in Sr and Zr isotopes than for example in Mo, Ru and Pd isotopes \cite{hot91}. Due to the scarcity of experimental information,
the gradual character of the deformation development transition in the Pd isotopes is
still under investigation. The calculations cited above \cite{ska97} predict that the spherical-deformed shape transition goes through a region of $\gamma$-softness in the chain of Pd isotopes. Indeed, the $E(4_1^+)/E(2_1^+)$ ratios for $^{108}$Pd, $^{110}$Pd and $^{112}$Pd are close to 
the gamma-soft $O(6)$ limit of 2.5. Furthermore, the energy staggering behavior of the $\gamma$-bands in the Pd and Ru isotopes was investigated 
in details in Ref. \cite{lal03}. It was concluded that $^{108,110,112}$Pd are softer than
$^{114,116}$Pd and that Pd isotopes are softer than Ru isotopes \cite{lal03}. The present study of $^{109}$Pd and $^{111}$Pd is
intended to shed more light on the question of $\gamma$-softness in the chain of Pd isotopes. Moreover, both nuclei
may be expected to form the region of maximum $\gamma$-softness for the Pd chain of isotopes.

Prior to the present work, a search for band structures in the neutron-rich nuclei $^{109}$Pd and $^{111}$Pd was performed using a fusion-fission reaction \cite{kut98} and the negative-parity $\nu(h_{11/2})$ bands were identified. The first crossing is proposed to be a result of $\nu(h_{11/2})^2$ alignment, which stabilizes the prolate shape \cite{kut98}. How to understand that in the light of the expected enhanced $\gamma$-softness? The situation was expressed in Ref. \cite{kru99} as "a prolate, but still $\gamma$-soft shape is stabilized with increased rotational frequency" in $^{110,112,114,116}$Pd isotopes. The basis of this assumption are cranked Woods-Saxon and total Routhian surface (TRS) calculations \cite{kru99}. With respect to the first band crossings in Pd isotopes heavier than $^{111}$Pd, which was argued to be due to a pair of $h_{11/2}$ neutrons or $g_{9/2}$ protons, the recent studies definitely agrees on $h_{11/2}^2$ neutrons alignment. Indeed, that is the case with the $h_{11/2}$ bands in $^{111,113,115}$Pd \cite{kut98,kru99,hou99,zha99} and with the ground-state bands in $^{112,114,116,118}$Pd \cite{lal03,hou99,zha01}.

Most of the nuclei in the discussed region are neutron rich and are not accessible through conventional fusion-evaporation 
in-beam techniques. Therefore, the experimental information is incomplete, even at low spin.
Before the present study, the information on the positive-parity states in $^{109}$Pd and $^{111}$Pd was limited to a very 
few low-energy levels with no connections in a band structure. 
Positive-parity bands built on the 5/2$^+$ ground states were observed in $^{101,103}$Pd \cite{ric73,nya99}, $^{105}$Pd \cite{ric77},
$^{107}$Pd \cite{poh96} as well as in $^{113,115}$Pd \cite{zha99}, but not in $^{109}$Pd and $^{111}$Pd. Filling this gap of experimental information may shed light on the deformation evolution of the positive-parity ground-state bands through the transitional region. Considering that this is expected to be a region of maximum $\gamma$-softness for the transitional Pd isotopes, the observation of these structures is important for understanding the whole transitional region.

The aim of the present work was to find positive-parity bands in $^{109}$Pd and $^{111}$Pd in order to fill up the
gap of missing experimental information. Furthermore, to investigate how these nuclei behave in the expected region of maximum
$\gamma$-softness from one hand and a prolate driving alignment of a pair of $h_{11/2}$ neutrons on the other hand.

\section{Experiment and analysis}
\label{sec:exp}
 
Excited states in the neutron-rich nuclei $^{109}$Pd and $^{111}$Pd were populated via an induced fusion-fission channel of the 
$^{30}$Si + $^{168}$Er reaction. Excited yrast states of $^{98,100,102}$Mo, produced also as fission fragments in the same experiment, were 
already studied and published \cite{lal07}. 
The beam of $^{30}$Si at an energy of 142 MeV was provided by the XTU tandem accelerator at the Legnaro National Laboratory. 
Prompt $\gamma$ rays emitted from the excited nuclei were detected with the EUROBALL III \cite{sim97} multidetector array 
consisted of 30 single HpGe Compton-shielded detectors, 26 Clover \cite{duc99} and 15 Cluster \cite{ebe90} composite Compton-shielded detectors. The $^{168}$Er target foil of 1.15 mg/cm$^2$ thickness was evaporated on a 9 mg/cm$^2$ 
gold backing, in which the recoiling nuclei were slowed down and finally stopped. Events were collected with the 
requirement that at least five unsuppressed Ge detectors fired in coincidence. A total of about 2$\cdot$10$^9$ 
three- and higher-fold events were recorded. In the off-line analysis two $E_{\gamma}$-$E_{\gamma}$-$E_{\gamma}$ cubes 
with energy ranges up to 4 MeV and up to 1 MeV, respectively, were sorted using
the RADWARE package \cite{rad95}. Triple gamma-ray coincidence
spectra were then produced by setting double gates on the  $E_{\gamma}-E_{\gamma}-E_{\gamma}$ cubes.

\begin{figure}  
  
\includegraphics[height=10.0cm, angle=-90]{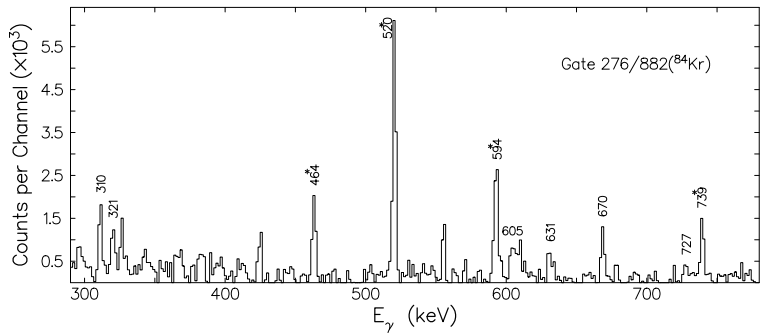}  
\caption{\label{fig:PdKr} Spectra, double gated on the 276 keV transition from $^{109}$Pd and the 882 keV transition from $^{84}$Kr.
Contaminant lines are marked with asterisks $(*)$.}
   
\end{figure}

In order to search for new transitions in $^{109}$Pd and $^{111}$Pd, populated as fission fragments, we examined spectra double gated on low-lying transitions in the complementary fission fragments $^{84,82}$Kr, respectively.
They clearly revealed the transitions of the  even-even  neighbors $^{108,110}$Pd in the $^{84}$Kr spectra and $^{110,112}$Pd in the $^{82}$Kr spectra, observed in the $6n$ and $4n$ channels, respectively. The transitions from the previously known negative-parity $\nu(h_{11/2})^2$ bands \cite{kut98} were seen in the respective $5n$ channels, confirming the observation of the odd Pd isotopes.
The structures built on the ground states in the odd-mass Pd isotopes are expected to be populated weaker than the yrast $\nu(h_{11/2})^2$ bands. Indeed, the data confirmed this expectation. In $^{109}$Pd, only the 276 keV transition and 597 level were known \cite{nds109Pd}. In order to identify the states above the known low-energy positive-parity levels, double gates on one of these transitions with the lowest-lying yrast transitions in the complementary Kr isotopes were created. A sample spectrum,
double gated on the 276 keV transition from $^{109}$Pd and on the 882 keV transition from $^{84}$Kr, is shown in Fig. \ref{fig:PdKr}.
Three transitions (230, 293, 523 keV) were observed previously \cite{nds111Pd} in $^{111}$Pd. In this case, a double gate on 230 and
293 keV transitions, known to be in coincidence, was used in addition to cross-double gates on the known transitions in $^{111}$Pd and
on the transitions of Kr isotopes to identify new transitions in $^{111}$Pd. Although the new transitions are weak and the spectra are contaminated, the coincidences were unambiguous and level schemes of the positive-parity structures in  $^{109,111}$Pd are constructed. 

The low statistics does not allow the deduction of relative $\gamma$-ray intensities with reasonable precision for
the observed transitions in $^{109,111}$Pd. In order to study the levels decay branches further experiments are needed.

The spin and parity assignments are based on the previously known adopted values for the lowest-lying
excited positive-parity states \cite{nds109Pd,nds111Pd}, systematics of the neighboring nuclei and on the fact that well 
defined band structures, with $\Delta I$ = 2 and $\Delta I$ = 1 transitions, are developed on top of them. Based on all that, the spin and parity assignments can be done unambiguously.

\section{Experimental results}

Positive-parity rotational (quasi-rotational) bands were observed for the first time in $^{109}$Pd and $^{111}$Pd in the
present work and are shown in Figs. \ref{fig:schemep109} and \ref{fig:schemep111}. Low-lying negative-parity yrast bands in 
both nuclei were previously observed in \cite{kut98} and are displayed in Fig. \ref{fig:schemenPd}. They are given
for completeness. 
  
\subsection{Positive-parity band in $^{109}$Pd}

\begin{figure}  
  
\includegraphics[height=10.0cm, angle=-90]{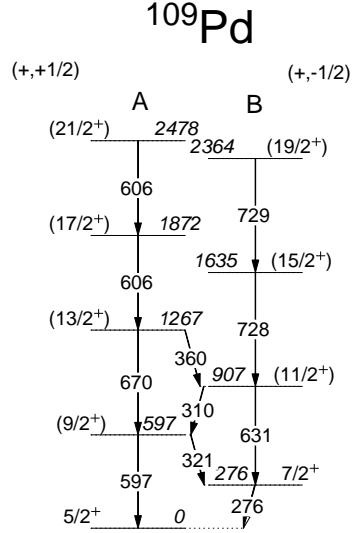}  
\caption{\label{fig:schemep109} The positive-parity ground-state band of $^{109}$Pd deduced 
from the present work. The energies are given in keV. The experimental uncertainties of the energies are less than 1 keV.
Only the 7/2$^+$ state at an energy of 276 keV with the 276 keV transition depopulating it to the ground state and the 597 keV level (the transition depopulating the 597 keV level was not observed) were observed in previous studies \cite{nds109Pd}.}
   
\end{figure}

No band structures built on the two known low-lying positive-parity states were observed in $^{109}$Pd before the present
study. The first 7/2$^+$ and (9/2$^+$) states at energies of 276 keV and 597 keV, respectively, were observed previously
as single not connected levels. A positive-parity band, with two signature partners developing on both these states, was 
observed in the present experiment as presented in Fig. \ref{fig:schemep109}.
The 7/2$^+$ state at 276 keV, as well as the one at 597 keV assumed to be with spin and parity of 7/2$^+$ or 9/2$^+$,
were firstly observed by M. Kanazawa et al. \cite{kan78} in $\beta$$^-$-decay experiment. The 7/2$^+$ state together with
the 276 keV transition depopulating it to the ground state
were also observed by Casten et al. \cite{cas80} in $(n,\gamma)$ experiment, while a 596(4) keV state was observed
in $(d,t)$ experiment \cite{die70} with an assignment of 7/2$^+$. The 597 keV transition was not observed before
the present study. We assign (9/2$^+$) to the 597 keV state based on the band structure observed.
Based on the fact that a rotational (quasi-rotational) band is observed, the previously existed experimental assignment of 7/2$^+$ for 
the 276 keV state as well as systematics of the ground-state bands in the region, the spin and parity of all levels can be proposed quite unambiguously. Three sample spectra double gated on the 276-321 keV, 276-728 keV and on the 310-728 keV transitions are shown in 
Fig. \ref{fig:spectra109}.

\subsection{Positive-parity band in $^{111}$Pd}   

\begin{figure}  
  
\includegraphics[height=10.0cm, angle=-90]{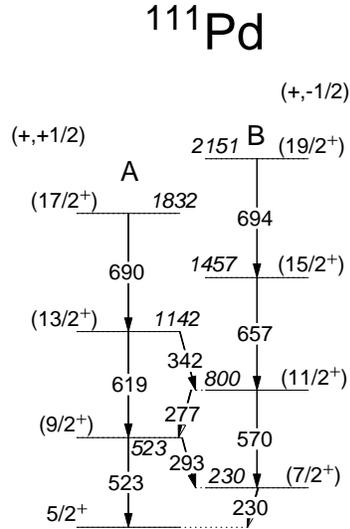}  
\caption{\label{fig:schemep111} The positive-parity ground-state band of $^{111}$Pd deduced 
from the present work. The energies are given in keV. The experimental uncertainties of the energies are less than 1 keV. 
The levels at 230 keV and 523 keV as well as the 230, 293 and 523 keV transitions were previously known \cite{nds111Pd}.}
  
\end{figure}

No positive-parity rotational band was observed in $^{111}$Pd before the present experimental study. 
Only two excited positive-parity states were known in $^{111}$Pd prior to the present study, but
not with the assigned spin and parity. A rotational band, with two signature partners developing on the two known levels, was
extracted from the present experimental data and presented in Fig. \ref{fig:schemep111}.
The state at 230 keV was previously observed in $(d,p)$, $\beta$$^-$-decay and $(n,\gamma)$ experiments
with possible spin and parity assignments of 7/2$^+$ or 9/2$^+$ \cite{nds111Pd}. The 230 keV transition depopulating it
to the ground state was also observed \cite{nds111Pd}. The level at 523 keV was observed before in 
$\beta$$^-$-decay experiments \cite{nds111Pd}. However, no spin assignment was adopted for it \cite{nds111Pd}.
The 293 keV and 523 keV transitions depopulating the state at 523 keV to the 230 keV level
and to the ground state, respectively, were also known \cite{nds111Pd}. 
Based on the observed rotational (quasi-rotational) band structure with $\Delta I$ = 1 and $\Delta I$ = 2 transitions and on the systematics of the ground-state bands in the region,
the spins and parities of the observed states can be assigned quite unambiguously.  
Two sample spectra double gated on the 230-570 keV transitions and on the 277-657 keV transitions are shown 
in Fig. \ref{fig:spectra111}.

\begin{figure}  
  
\includegraphics[height=14.0cm, angle=-90]{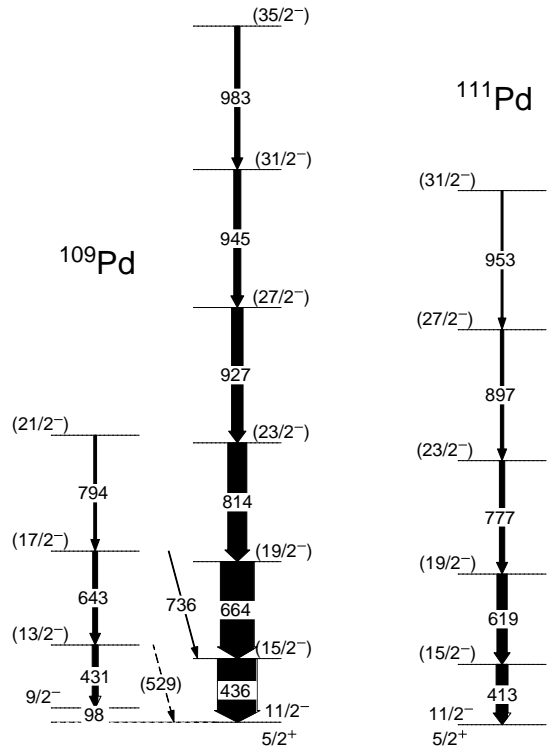}  
\caption{\label{fig:schemenPd} Level scheme of $^{109}$Pd and $^{111}$Pd for which only negative-parity bands were known so far
\cite{kut98}.}
  
\end{figure}  

\begin{figure}  
  
\includegraphics[height=14.0cm]{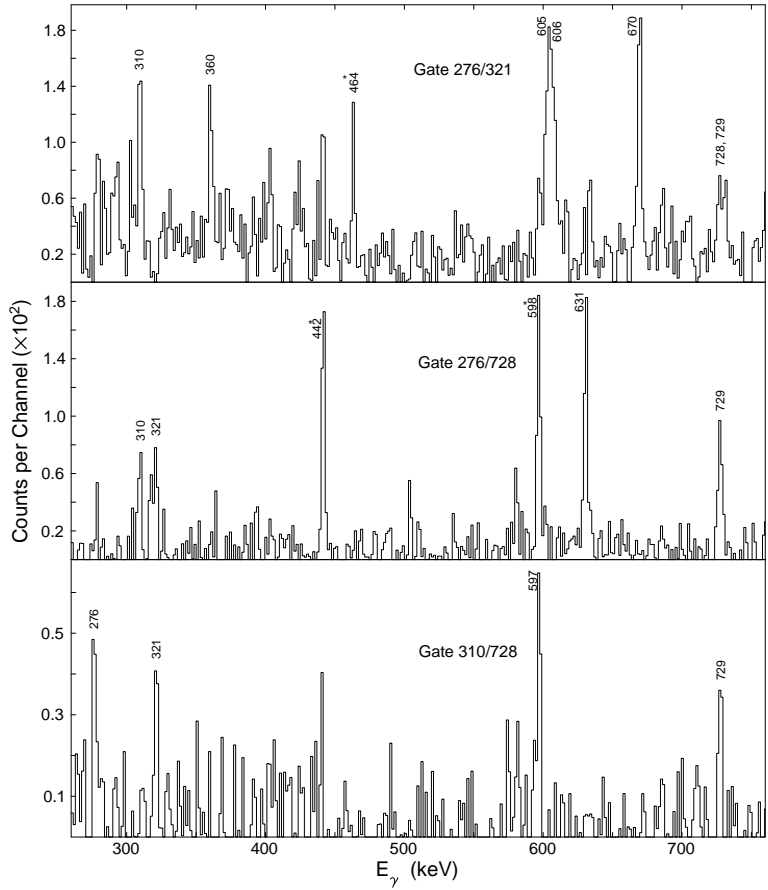}  
\caption{\label{fig:spectra109} Sample spectra double gated on the 276-321 keV, 276-728 keV and on the 310-728 keV transitions in 
$^{109}$Pd. The lines marked with asterisks $(*)$ are from contaminants.}
  
\end{figure}  

\begin{figure}  
  
\includegraphics[height=14.0cm, angle=-90]{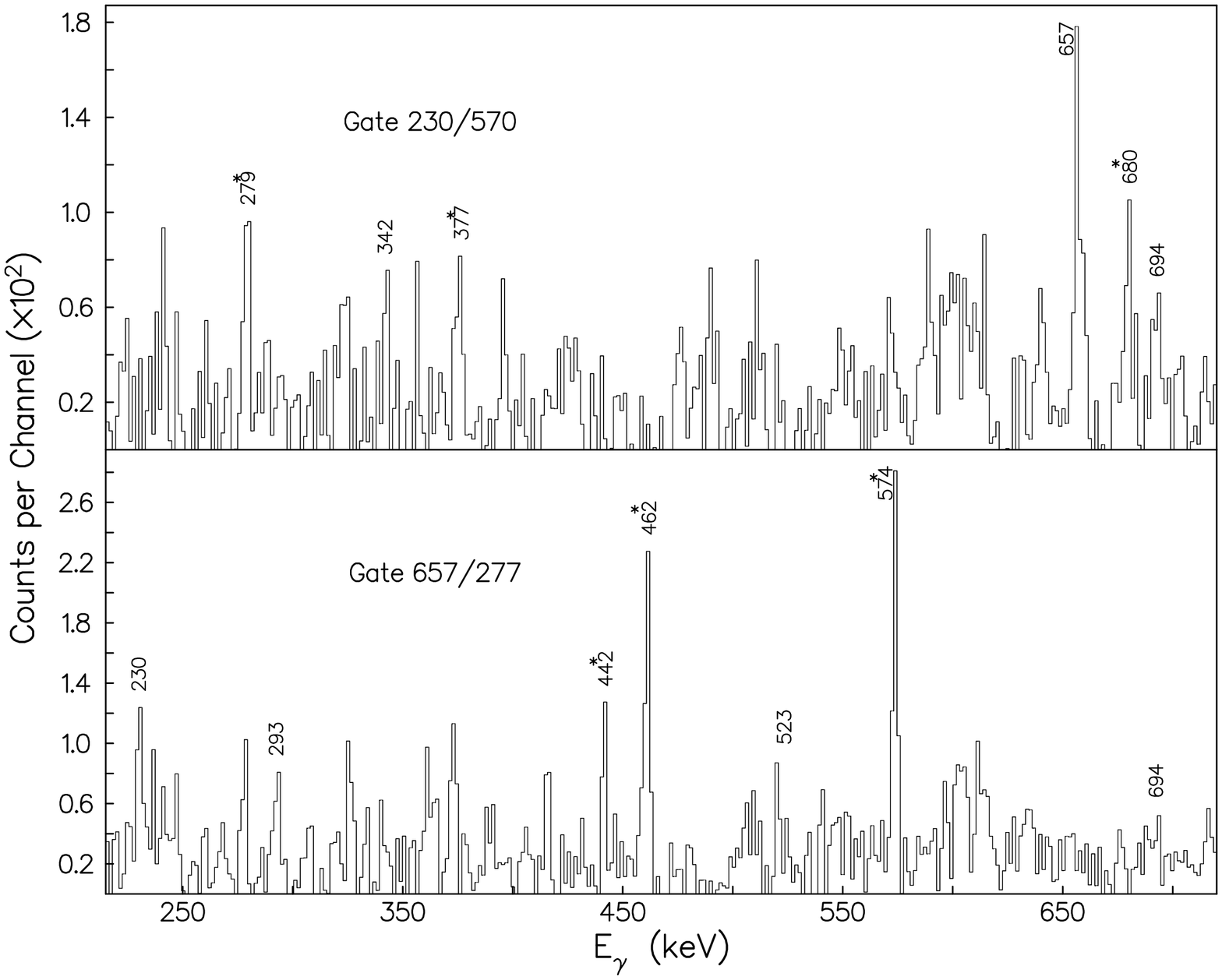}  
\caption{\label{fig:spectra111} Sample spectra double gated on the 230-570 keV and on the 277-657 keV transitions in $^{111}$Pd.
The lines marked with asterisks $(*)$ are from contaminants.}
  
\end{figure}

\section{Discussion}

\subsection{A region of $\gamma$-softness}

The nuclei $^{109}$Pd and $^{111}$Pd lie in the transitional region between vibrational and rotational nuclei.
It is known that this transition from sphericity to deformation in the Pd isotopes is more gradual than it is in the Sr and Zr 
isotopes \cite{hot91}, which form one of the border at $N =$ 50 and $Z =$ 38,40 of the neutron-rich $A \approx$ 100 region. 
As it is also studied \cite{ska97}, this transition in palladium goes rather through $\gamma$-softness than just through softness 
to axial quadrupole deformation. Taking into account that it is not straight forward to distinguish between triaxiality and 
$\gamma$-softness experimentally, the appearance of stable triaxiality is also possible. 
The ratios of $E(4_1^+)/E(2_1^+)$ is 2.42 for $^{108}$Pd \cite{nds108Pd} and 2.46 for $^{110}$Pd \cite{nds108Pd}. These values are very 
close to the ones predicted by the IBM-2 model value of 2.5 for a $\gamma$-soft $O(6)$ rotor. 
Additional arguments may come from the investigation of the $\gamma$-bands in the region. 
In principle, an odd-even energy-level staggering in a $\gamma$-band is known to be a sign of a
non-axial shape (soft or rigid) \cite{cas00}. The staggering effect has been analyzed in a framework of a ground-state band and a 
$\gamma$-band interaction (see for example \cite{min00} and references therein). 
Such an energy staggering  was actually observed. The levels of the $\gamma$-bands in
$^{108}$Pd \cite{lal03}, $^{110}$Pd \cite{lal03}, $^{112}$Pd \cite{kru01} are grouped as 2$^+$, (3$^+$,4$^+$), (5$^+$,6$^+$) and etc., which 
is consistent with $\gamma$-soft behavior accordingly to the model of Wilets and Jean \cite{wil56}. On the contrary, the Davidov's model 
describing rigid triaxiality predicts the levels of the $\gamma$-band to be grouped as (2$^+$,3$^+$), (4$^+$,5$^+$) and etc. \cite{dav58}. 
The 2$^+$, (3$^+$,4$^+$), (5$^+$,6$^+$) spacing has its maximum expression in the Pd isotopic chain in $^{110}$Pd, while in the $\gamma$-bands 
of the even-even isotopes lighter than $^{108}$Pd and heavier than $^{112}$Pd, the level spacing is going towards more equidistant or 
regular distribution. Such behaviors suggest more axially symmetric quasi-rotations (with quadrupole vibrational mixing) from one 
hand or more rigid axial symmetry rotations on the other hand. 
 
The $\gamma$-bands in the transitional Pd region were systematically discussed in \cite{lal03} and compared with the
$\gamma$-bands systematics in corresponding Ru isotones. Similarly, $\gamma$-softness rather than rigid triaxiality was
suggested. The staggering effect was studied. It revealed that the energy staggering in $^{108}$Pd, $^{110}$Pd and $^{112}$Pd
has a larger amplitude that in the neighboring Pd isotopes. While the $\gamma$-bands are not so well developed in $^{102,104,106}$Pd as 
in the heavier Pd isotopes, the amplitude of the  staggering effect decreases in $^{114}$Pd and $^{116}$Pd \cite{lal03}. It was concluded that $^{114,116}$Pd are not so soft as $^{108}$Pd, $^{110}$Pd and $^{112}$Pd as well as that the Pd isotopes are softer than Ru isotopes \cite{lal03}.  
The experimentally observed decrease in the energy of the 2$^+$ states (see Fig. 6 in \cite{lal03}) of the $\gamma$-bands from 1128 keV in 
$^{106}$Pd (N=60) to 695 keV in $^{114}$Pd (N=68) was suggested to be due to an increase of collectivity \cite{lal03}. Beyond $^{114}$Pd, 
the energy of the 2$_2^+$ state increases again. Similar is the behavior of the 2$_1^+$ states, which energies decrease up to $^{114}$Pd 
($N=68$) and then increase again. The ratio $E4_1^+/E2_1^+$ also reaches its maximum of 2.6 at $N = 68$. The behavior of $^{114}$Pd may 
have midshell effects origin as $N = 68$ is approximately between $N$ = 50 and $N$ = 82.

In agreement with the above discussion based on experimental observables, the Pd isotopes are also predicted to be $\gamma$-soft at 
ground state by the Nilsson-Strutinski calculations \cite{ska97}.
In Ref. \cite{kru99}, TRS calculations using Woods-Saxon potential as well as Ultimate Cranker were performed
for the even-even Pd isotopes from $^{110}$Pd to $^{116}$Pd. Both type of calculations predict $\gamma$-softness at low-rotational frequencies.

Thus, one may expect a significant degree of $\gamma$-softness at low energies and angular momenta for the nuclei of interest,
$^{109}$Pd and $^{111}$Pd. Moreover, according to the behavior of the $\gamma$-bands in the even-even Pd isotopes, one may suggest that both nuclei 
lie in the region of maximum $\gamma$-softness. This suggestion is also supported by the TRS calculations \cite{wys88} reported below.
  
\subsection{Negative-parity bands in $^{109}$Pd and $^{111}$Pd}
 
The only rotational structures in $^{109}$Pd and $^{111}$Pd observed before the present study were the
negative-parity bands built on the low-lying 11/2$^-$ isomer \cite{kut98} (see Fig. \ref{fig:schemenPd}). They 
were interpreted as built on an orbital in the middle of $h_{11/2}$ neutron subshell with a prolate deformation \cite{kut98}.
Analogous bands were observed in the neighboring odd-mass Pd isotopes as for example in $^{105}$Pd \cite{ric77}, $^{107}$Pd \cite{poh96},
$^{113}$Pd \cite{kru99,hou99,zha99,fon05}, $^{115}$Pd \cite{kru99,zha99,hou99,fon05}. They appear as quasi-rotationally aligned except for 
the ones in $^{109}$Pd \cite{kut98} and $^{113}$Pd \cite{hou99}, where the signature partner was observed. The $\Delta I$ = 1 staggering 
of the level energies of the signature partners suggests kind of semi-decoupled or so called Coriolis-distorted structures. 
It has not been possible so far, to investigate on the origin of the different band behavior observed
in $^{109}$Pd and $^{113}$Pd and in the other odd-even isotopes. 
It could be that the second partners were just not observed in the other isotopes or that they appear, forming a $\Delta I$ = 1
band staggering pattern, only in the region of maximum $\gamma$-softness. In the latter case, as $^{111}$Pd lies
in this region of maximum $\gamma$-softness, the second signature partner, energetically shifted towards the first one, 
may be expected here as well. An attempt to search for it in our data was unsuccessful, probably due to the low statistics. 
 
The $h_{11/2}$ bands in the odd-mass Pd isotopes from $^{107}$Pd to $^{115}$Pd show the first band crossing at a frequency of 
about 0.47 MeV \cite{poh96,kut98,hou99,kru99,zha99}, suggesting similar crossing nature for all of them.
It is at about 0.56 MeV in $^{105}$Pd \cite{ric77} explained as a result of its smaller deformation \cite{ric77,poh96}. 
The crossing in the odd-mass isotopes is delayed with respect to the first crossing ($\approx$ 0.35 MeV) in the neighboring 
even-mass Pd isotopes, explained to be due to a blocking effect caused by the $h_{11/2}$ neutron.
Therefore, the alignment of $h^2_{11/2}$ neutrons rather than $g^2_{9/2}$ protons, which was the alternative interpretation,
was proposed to be the nature of this crossing in the odd-mass Pd isotopes (see e.g. \cite{poh96,kut98,hou99,kru99,zha99}) and in the 
even-mass Pd isotopes $^{108}$Pd \cite{poh96,reg97,lal03}, $^{110}$Pd \cite{hou99,lal03}, $^{112,114,116,118}$Pd \cite{hou99,zha01}. 
Additional arguments supporting the $h^2_{11/2}$ neutron alignment comes for example from CSM calculations \cite{reg97} and 
TRS calculations \cite{poh96} performed for $^{108}$Pd, and cranked-HFB for $^{112}$Pd \cite{hou99}. 

It is known that the alignment of $h^2_{11/2}$ neutrons drives the nucleus towards prolate shape. Indeed, in the
mentioned above calculations, prolate driving force is predicted by the CSM and cranked-HFB models for the aligned pair of $h_{11/2}$ 
neutrons. At the same time, Nilsson-Strutinski calculations \cite{ska97} predict very $\gamma$-soft potential for the transitional Pd 
nuclei. The TRS calculations for $^{108}$Pd \cite{poh96} 
also show a shallow prolate minimum, soft with respect to $\gamma$- and $\beta$-deformation at $\hbar w =$ 0 MeV. With an increase of the 
rotational frequency the minimum moves towards soft non-axial deformation.
Then, the alignment of $h_{11/2}$ pair of neutrons stabilizes the prolate shape, but the potential surface is still quite soft towards $\beta$- and $\gamma$-deformation \cite{poh96}. The performed TRS calculations for $^{110}$Pd (not shown) revealed that the potential is more $\gamma$-soft 
than that in $^{108}$Pd. Kruecken et al. \cite{kru99}, reported on TRS calculations for the even-even isotopes $^{110,112,114,116}$Pd. They 
predict very $\gamma$-soft minima at low rotational frequency. After the alignment of the $h_{11/2}$ pair
of neutrons, the prolate but otherwise $\gamma$-soft shape is going towards stabilization \cite{kru99}. Will a shallow prolate minimum of a very $\gamma$-soft potential be the case of $^{109}$Pd and $^{111}$Pd as well?

\begin{figure} 

\includegraphics[height=16.0cm]{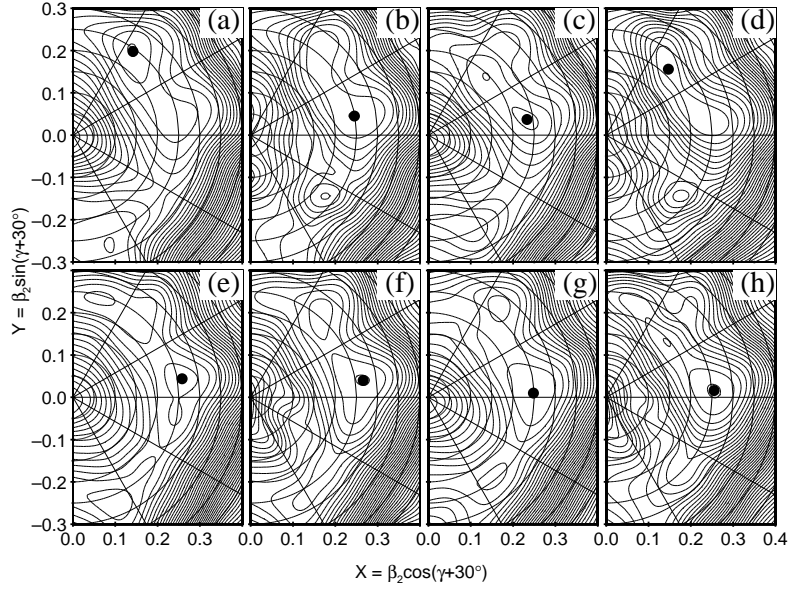}  
\caption{\label{fig:TRSn} Total Routhian surface calculations for the negative-parity bands in $^{109}$Pd and $^{111}$Pd. Contour 
lines are in 200 keV increment. (a) $^{109}$Pd, (-,+1/2), $\hbar w$ = 0.100 MeV, $\beta_2$ = 0.243, $\gamma$ = +24$^{\circ}$; 
(b) $^{109}$Pd, (-,+1/2), $\hbar w$ = 0.400 MeV, $\beta_2$ = 0.249, $\gamma$ = -20$^{\circ}$; (c) $^{109}$Pd, (-,-1/2), $\hbar w$ = 0.100 MeV, 
$\beta_2$ = 0.236, $\gamma$ = -21$^{\circ}$; (d) $^{109}$Pd, (-,-1/2), $\hbar w$ = 0.400 MeV, $\beta_2$ = 0.215, $\gamma$ = +17$^{\circ}$; 
(e) $^{111}$Pd, (-,+1/2), $\hbar w$ = 0.100 MeV, $\beta_2$ = 0.262, $\gamma$ = -20$^{\circ}$; (f) $^{111}$Pd, (-,+1/2), $\hbar w$ = 0.350 MeV, 
$\beta_2$ = 0.269, $\gamma$ = -21$^{\circ}$; (g) $^{111}$Pd, (-,-1/2), $\hbar w$ = 0.100 MeV, $\beta_2$ = 0.248, $\gamma$ = -28$^{\circ}$; 
(h) $^{111}$Pd, (-,-1/2), $\hbar w$ = 0.350 MeV, $\beta_2$ = 0.255, $\gamma$ = -26$^{\circ}$.}
  
\end{figure}

In Fig. \ref{fig:TRSn}, TRS calculations \cite{wys88} for the negative-parity bands built on the 11/2$^-$ states in $^{109}$Pd and $^{111}$Pd 
are shown. The calculations were performed separately for both signatures in both nuclei. At $\hbar w =$ 0.100 MeV, the $\alpha$ = +1/2 
signature partner in $^{109}$Pd is calculated with a shallow minimum at $\gamma$ = +24$^{\circ}$, while the $\alpha$ = -1/2 signature partner 
is at $\gamma$ = -21$^{\circ}$. Both surfaces reveal pronounced $\gamma$- and $\beta$-softness. The described pictures persist in both nuclei 
up to around the frequency at which the crossing is expected. It must be noted that both minima lie within a shallow $\gamma$-soft valley 
between $\gamma$ $\approx$ +30$^{\circ}$ to $\gamma$ $\approx$ -30$^{\circ}$. Interpreting the calculations one shall consider that the minima 
are very shallow and the energy potential surface is predicted very soft and they may be very sensitive to the calculation parameters.
Indeed, the average $\gamma$-deformation of both minima is $\gamma$ $\approx$ 0$^{\circ}$, which corresponds to a prolate shape.
On the other hand, a non-axial very $\gamma$-soft shape is predicted for the branches itself. This behavior can be described
as two different shallow potential wells almost symmetrically displaced from $\gamma$ = 0$^{\circ}$ for both signatures. The very low barrier 
between them would easily allow a change of the $\gamma$-deformation. Although only $\alpha$ = -1/2 signature
was observed in $^{111}$Pd, calculations for both band signatures are presented in Fig. \ref{fig:TRSn}. A pronounced $\gamma$-deformation is 
predicted for both signatures and, especially for $\alpha$ = -1/2 one, at $\hbar w =$ 0.100 MeV with $\gamma$ = -28$^{\circ}$ and $\beta_2$ = 0.248. Independently of the softness of the potential with very shallow minima and of the uncertainties of the present
calculations, the TRS calculations reveal a different picture for $^{109}$Pd and $^{111}$Pd than for $^{107}$Pd for example. While the
TRS calculations predict $\gamma$-soft prolate minima for this band in $^{107}$Pd (not presented), 
$^{109}$Pd and $^{111}$Pd are predicted quite $\gamma$-soft with non-axial minima. 

The TRS calculations do predict a band crossing with $h^2_{11/2}$ neutrons in agreement with the above discussion.
The alignment of the $h_{11/2}$ pair of neutrons in $^{109}$Pd moves the minimum in the $\alpha$ = +1/2 signature branch 
from $\gamma$ = +24$^{\circ}$ to $\gamma$ = -20$^{\circ}$, while the opposite (minimum moves from $\gamma$ = -21$^{\circ}$ 
to $\gamma$ = +17$^{\circ}$) is
predicted for the $\alpha$ = -1/2 signature. The potential is still very $\gamma$-soft.
In $^{111}$Pd, on the other hand, the $\gamma$ $\approx$ -28$^{\circ}$ minimum of the $\alpha =$ -1/2 signature persists through 
the alignment of the $h_{11/2}^2$ neutrons. The TRS calculations predict both signatures in $^{111}$Pd with similar $\gamma$-deformation, 
less $\gamma$-softness and a stronger $\beta_2$ deformation due to the alignment of $h_{11/2}^2$ neutrons. Thus, the alignment of 
$h_{11/2}^2$ neutrons stabilizes the triaxial minimum at negative $\gamma$-deformation in $^{111}$Pd. Prolate and not so soft shapes 
are calculated for both nuclei at higher frequency ($\hbar w =$ 0.600 MeV)  and angular momentum.

\subsection{Positive-parity bands in $^{109}$Pd and $^{111}$Pd}

In the present experiment, the ground-state positive-parity bands were observed
in $^{109}$Pd and in $^{111}$Pd. They exhibit energy-staggered $\Delta I$ = 1 transitions and thus, appear as semi-decoupled bands.
Reasons for an energy-staggered appearing of a band can be due to a band mixing, Coriolis distortion due to high-$j$ orbital, $\gamma$-softness, triaxiality. The observed bands pattern may be explained by the $\gamma$-softness of $^{109}$Pd and $^{111}$Pd.
However, analogous bands, on top of the 5/2$^+$ ground state with two energetically-staggered signature partners, were also observed
in $^{101}$Pd \cite{ric73}, $^{103}$Pd \cite{nya99}, $^{105}$Pd \cite{ric77}, $^{113}$Pd \cite{zha99}, $^{115}$Pd \cite{zha99}. The $\Delta I$ = 1 staggering of this band in $^{101}$Pd \cite{ric73} for example can not be explained by $\gamma$-softness as this nucleus is expected to be more 
$\beta$-soft rather than $\gamma$-soft. It may be rather due to a band mixing for instance. It can be seen from the level schemes of these nuclei
 that bands or levels which may cause band mixing have, indeed, been observed. In the heavier $^{113}$Pd, on the other hand, the positive-parity 
band may be explained by $\gamma$-softness. Again, we have performed TRS calculations for the ground state of $^{113}$Pd and they clearly show this 
nucleus being considerably $\gamma$-soft. In $^{107}$Pd \cite{poh96}, on the other hand, a strongly populated band (band 4) looking like a quasi-rotationally aligned was observed. The TRS calculations for this band (not shown) reveal a $\gamma$-soft prolate minimum for the $\alpha$ = -1/2 signature and a nearly prolate $\gamma$-soft minimum with $\gamma =$ 10$^{\circ}$ for the $\alpha =$ +1/2 signature. Thus, the calculations predict a signature splitting. A more detailed theoretical study is needed in order to explain the behavior of the positive-parity ground-state bands in  
the chain of the odd-A Pd isotopes.
 
\begin{figure} 

\includegraphics[height=16.0cm]{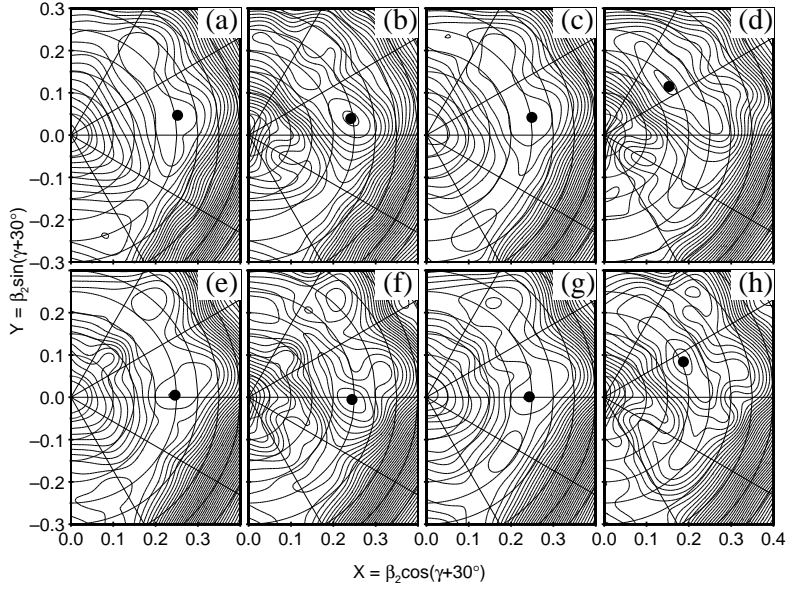}  
\caption{\label{fig:TRSp} Total Routhian surface calculations for the positive-parity bands in $^{109}$Pd and $^{111}$Pd. Contour
lines are in 200 keV increment. (a) $^{109}$Pd, (+,+1/2), $\hbar w$ = 0.100 MeV, $\beta_2$ = 0.256, $\gamma$ = -19$^{\circ}$; 
(b) $^{109}$Pd, (+,+1/2), $\hbar w$ = 0.350 MeV, $\beta_2$ = 0.245, $\gamma$ = -21$^{\circ}$; (c) $^{109}$Pd, (+,-1/2), $\hbar w$ = 0.100 MeV, 
$\beta_2$ = 0.253, $\gamma$ = -20$^{\circ}$; (d) $^{109}$Pd, (+,-1/2), $\hbar w$ = 0.350 MeV, $\beta_2$ = 0.191, $\gamma$ = +7$^{\circ}$; 
(e) $^{111}$Pd, (+,+1/2), $\hbar w$ = 0.100 MeV, $\beta_2$ = 0.246, $\gamma$ = -29$^{\circ}$; (f) $^{111}$Pd, (+,+1/2), $\hbar w$ = 0.350 MeV, 
$\beta_2$ = 0.244, $\gamma$ = -31$^{\circ}$; (g) $^{111}$Pd, (+,-1/2), $\hbar w$ = 0.100 MeV, $\beta_2$ = 0.244, $\gamma$ = -30$^{\circ}$; 
(h) $^{111}$Pd, (+,-1/2), $\hbar w$ = 0.350 MeV, $\beta_2$ = 0.204, $\gamma$ = -9$^{\circ}$.}
  
\end{figure}

While $^{107}$Pd is predicted to be moderately $\gamma$-soft at its ground state, but still around prolate deformation by the TRS calculations, the TRS plots presented in Fig. \ref{fig:TRSp} for $^{109}$Pd and $^{111}$Pd draw a long valley from $\gamma =$ 60$^{\circ}$ till $\gamma =$ -120$^{\circ}$ revealing extreme $\gamma$-softness (especially for $^{111}$Pd). Indeed, the nucleus $^{109}$Pd at ground state looks considerably $\gamma$-soft with a minimum at $\gamma = -20^{\circ}$ and extreme $\gamma$-softness with a minimum at $\gamma = -30^{\circ}$ is seen for $^{111}$Pd. The corresponding predominant configurations are calculated to be $d_{5/2}$ neutrons for $^{109}$Pd and $g_{7/2}$ neutrons for $^{111}$Pd. These minima in both 
nuclei persist up to around the crossing with $h_{11/2}$ pair of neutrons.
Calculations were performed separately for both signature partners of these bands in $^{109}$Pd and $^{111}$Pd as shown in Fig. \ref{fig:TRSp}. The signature partners have almost the same deformation up to the crossing. 

\begin{figure} 

\includegraphics[height=10.0cm]{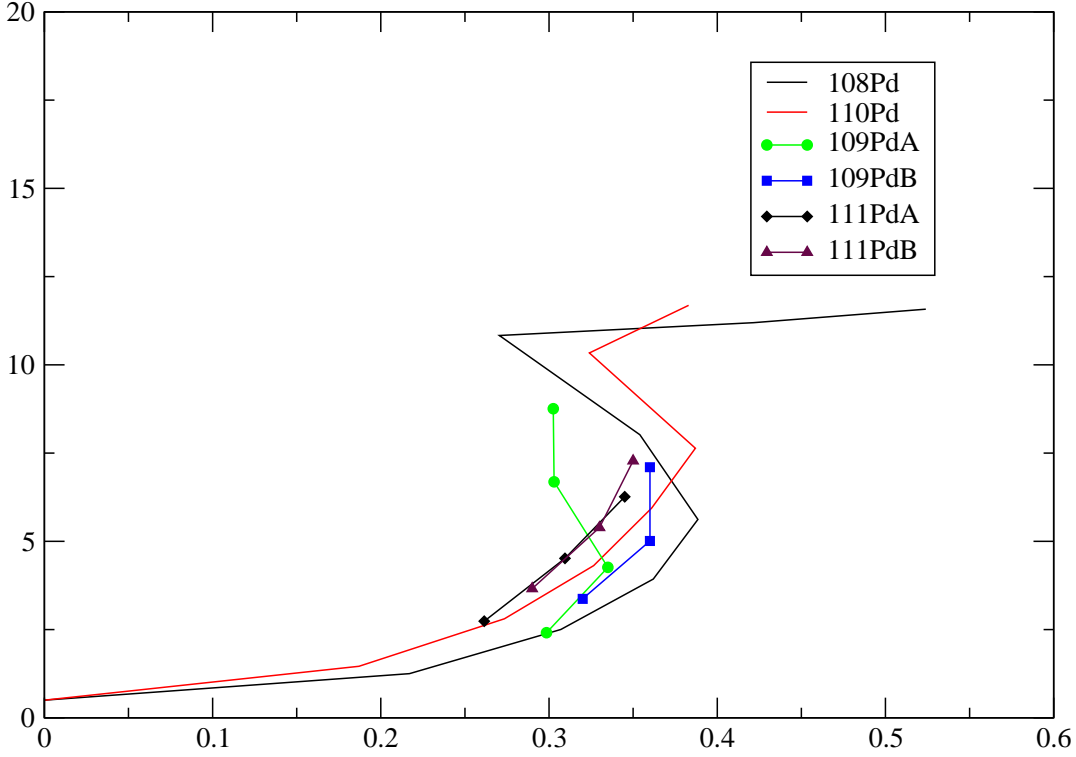}
\caption{\label{fig:alignment} (color online) Experimental alignments $i_x$ for the ground-state positive-parity bands in $^{109}$Pd 
and $^{111}$Pd and compared to the $i_x$ for the ground-state positive-parity bands in the neighboring even-even $^{108}$Pd and 
$^{110}$Pd isotopes. The Harris parameters used were $J_0$ = 5$\hbar^2$/MeV and $J_1$ = 16$\hbar^4$/MeV$^3$. For the bands in 
$^{109}$Pd and $^{111}$Pd, $K$ = 5/2 was used. The letters A, B correspond to the band signature partners labels used on the 
level schemes.}
  
\end{figure}

The alignments, $i_x$, of the newly observed positive-parity bands in $^{109}$Pd and $^{111}$Pd and for those in the
neighboring isotopes $^{108}$Pd and $^{110}$Pd are presented in Fig. \ref{fig:alignment}. The staggering pattern of the levels within the 
bands suggest that $K$ changes within the bands. However, the behavior of the $i_x$ does not change strongly with $K$ and $K =$ 5/2 is used 
for both nuclei. It is seen that the crossing in $^{109}$Pd occurs at around 0.35 MeV, which is at about the same frequency as in the 
even-even Pd neighbors. This behavior is consistent with an alignment of a pair of $h_{11/2}$ neutrons, as no blocking is expected 
in these bands. It cannot be seen in $^{111}$Pd in the present data. Similarly, the first crossings in the bands on top of the 
5/2$^+$ ground states in $^{107}$Pd \cite{poh96} and $^{113}$Pd \cite{zha99} are observed at $\hbar w$ $\approx$ 0.32 MeV. 
In both nuclei, the crossing was interpreted as caused by an alignment of $h^2_{11/2}$ neutrons. In agreement, an alignment 
of $h_{11/2}$ pair of neutrons is predicted by the TRS calculations for the positive-parity bands in $^{109}$Pd and $^{111}$Pd. 
The calculations predict that the alignment of the 
$h_{11/2}$ neutron pair (at frequency about 0.300 MeV) forces the $\alpha$ = -1/2 signature partners in $^{109}$Pd and $^{111}$Pd 
 towards more $\gamma$-stable near prolate shape. However, the $\alpha$ = +1/2 signature partners in both nuclei are predicted to 
preserve and actually to stabilize the triaxial deformation. Thus, a more stable triaxiality is predicted at higher spin. 
Experimental observation of the bands at higher spins in future experiments may allow to test this predictions. 
A negative-parity band with two signature partners connected with staggered $M1$ transitions was observed at higher 
excitation energies in $^{112}$Pd \cite{kru01}. TRS calculations also predict possibly a triaxial shape for this band \cite{kru01}.

Thus, basically all experimental observables, calculations and arguments suggest that $^{109}$Pd and $^{111}$Pd, probably 
together with $^{113}$Pd, lie in the region of maximum $\gamma$-softness for the transitional region of Pd isotopes. The alignment 
of $h_{11/2}$ pair of neutrons is predicted by the TRS calculations to drive the nuclei towards more stable deformation, 
prolate or even triaxial.

\section{Summary}

The neutron-rich nuclei $^{109}$Pd and $^{111}$Pd were produced as fission fragments at high-excitation energies and angular momentum
through the induced fusion-fission reaction $^{30}$Si + $^{168}$Er at a beam energy of 142 MeV. The beam of $^{30}$Si was provided 
by the XTU tandem accelerator at the Legnaro National Laboratory. Although few low-lying positive-parity states were known before
the present study, no band structures built on them were observed. Positive-parity bands in $^{109}$Pd and $^{111}$Pd were observed for 
the first time in this work and reasonable spin and parities assignments could be made.

It was concluded that $^{109}$Pd and $^{111}$Pd, together with $^{113}$Pd, lie in the transitional Pd region where maximum 
$\gamma$-softness is expected to occur. Indeed, the nuclei $^{109}$Pd and $^{111}$Pd are predicted to be $\gamma$-soft with 
shallow triaxial minima for the lowest-lying states of the observed negative-parity as well as positive-parity bands by the 
TRS calculations. An extreme $\gamma$-softness is predicted for $^{111}$Pd. The TRS calculations do confirm the proposed 
alignment of $h_{11/2}^2$ neutrons \cite{kut98} for the negative-parity bands built on the $11/2^-$ states in both nuclei. 
However, the stabilization towards prolate shape is predicted at higher frequencies than the crossing, while the crossing 
itself goes through $\gamma$-soft non-axial behavior.
The observed positive-parity bands in $^{109}$Pd and $^{111}$Pd appear like semi-decoupled bands, with a $\Delta I$ = 1 level-energy 
staggering, which can be consistent with a $\gamma$-soft potential \cite{cas00}.
The analogous band in $^{107}$Pd has a rotationally-aligned behavior while odd-even level-energy staggering is revealed for the 
analogous bands in lighter Pd isotopes ($^{101-105}$Pd). Detailed theoretical studies are needed in order to explain the 
behavior of the positive-parity ground-state bands in the chain of Pd isotopes. 
The first band crossing in the observed positive-parity bands of this work, is proposed to be caused by the alignment
of the $h_{11/2}^2$ neutrons. The calculations predict that this alignment drives one of the band signatures
in both nuclei to a less $\gamma$-soft near-prolate shape, while a stabilization of the triaxial shape is predicted for the other signature
in both nuclei. Further experiments are definitely needed in order to investigate on the
predicted band signature splitting and also search for a possible stable triaxiality.

\begin{acknowledgments}  

The authors would like to acknowledge the useful discussions with N. Minkov.
This work was partly supported from the Bulgarian National Science Fund, contract No: DMU02/1
and the collaboration agreement between Bulgarian Academy of Sciences-CNRS under contract No. 4847,
and by the German BMBF under Contract No. 06BN109.

\end{acknowledgments}  
  
\bibliography{tf.bib}  

\end{document}